\documentclass[prl,twocolumn,groupedaddress,amssymb,prb,aps,showpacs]{revtex4}
\usepackage{graphicx,bm,amsmath}
\begin{document}

\title{Superconducting qubit storage and entanglement with nanomechanical resonators}

\author{A.~N. Cleland$^1$ and M.~R. Geller$^2$}

\affiliation{$^1$Department of Physics, University of California, Santa Barbara, California 93106 \\
$^2$Department of Physics and Astronomy, University of Georgia,  Athens, Georgia 30602-2451}

\date{\today}

\begin{abstract}
We describe a quantum computational architecture based on integrating nanomechanical resonators
with Josephson junction phase qubits, with which we implement single- and multi-qubit operations.
The nanomechanical resonator is a GHz-frequency, high-quality-factor dilatational resonator,
coupled to the Josephson phase through a piezoelectric interaction. This system is analogous to one
or more few-level atoms (the Josephson qubits) in a tunable electromagnetic cavity (the
nanomechanical resonator). Our architecture combines the best features of solid-state and
cavity-QED approaches, and may make possible multi-qubit processing in a scalable, solid-state
environment.
\end{abstract}


\maketitle

The lack of scalable qubit architectures, with sufficiently long quantum-coherence lifetimes and a
suitably controllable entanglement scheme, remains the principal roadblock to building a
large-scale quantum computer. Superconducting devices exhibit robust macroscopic quantum behavior
\cite{Makhlin2001}. Recently, there have been exciting demonstrations of long-lived Rabi
oscillations in current-biased Josephson junctions \cite{Yu02,Martinis2002}, subsequently combined
with a two-qubit coupling scheme \cite{Wellstood:2003}, and in parallel, demonstrations of Rabi
oscillations and Ramsey fringes in a Cooper-pair box \cite{Nakamura1999,Na02,Vion2002}. These
accomplishments have generated significant interest in the potential for Josephson-junction-based
quantum computation \cite{Le02}. Coherence times $\tau_{\rm \varphi}$ up to 5$\, \mu$s have been
reported in the current-biased devices \cite{Yu02}, with corresponding quality factors $Q_{\rm
\varphi} \equiv \tau_{\rm \varphi} \, \Delta E / h$ of the order of $10^5$, yielding sufficient
coherence to perform many logical operations. Here $\Delta E$ is the qubit energy level spacing.

In this paper, we describe an architecture in which ultrahigh-frequency resonators coherently
couple two or more current-biased Josephson junctions, where the superconducting ``phase qubits''
are formed from the energy eigenstates of the junctions. We show that the system is analogous to
one or more few-level atoms (the Josephson junctions) in a tunable electromagnetic cavity (the
resonator), except that here we can individually tune the energy level spacing of each atom, and
control the electromagnetic interaction strength.

Other investigators have proposed the use of electromagnetic
\cite{Shnirman1997,Makhlin1999,Mooij1999,Makhlin2000,You2002,Plastina2003,Blais2002,Smirnov2002u}
or superconducting \cite{Buisson2001,Marquardt2001} resonators to couple Josephson junctions
together. The use of nanomechanical resonators to mediate multi-qubit operations has not to our
knowledge been described previously, although an approach to create entangled states of a single
resonator has been proposed \cite{Armour2002a}. The use of mechanical as opposed to electromagnetic
resonators has the advantage that potentially much higher quality factors can be achieved
\cite{Yang:2000}, with significantly smaller dimensions, enabling a truly scalable approach.

\begin{figure}
    \includegraphics[width=.4\textwidth]{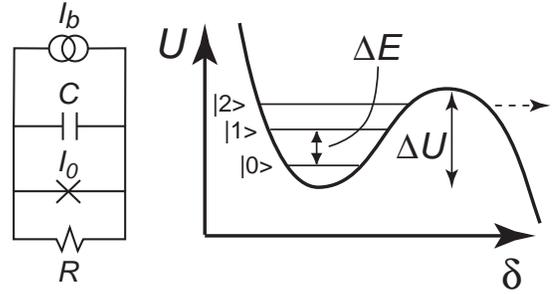}
    \caption[]{\label{fig.cubicpotential}\emph{Left:} Equivalent-circuit model for a current-biased
    Josephson junction. A capacitance $C$ and resistance $R$ in parallel with an ideal
    Josephson element with critical current $I_0$, all with a bias current $I_b$.
    \emph{Right:} Metastable potential well in the cubic potential limit,
    showing the barrier height $\Delta U$ that separates the metastable states from the
    continuum. Here there are three
    quasi-bound states $| 0 \rangle$, $| 1 \rangle$, and $| 2 \rangle$, the lower two
    separated  in energy by $\Delta E$.}
\end{figure}

Our implementation uses large-area current-biased Josephson junctions, with capacitance $C$ and
critical current $I_0$; a circuit model is shown in Fig. \ref{fig.cubicpotential}. The largest
relevant energy is the Josephson energy $E_{\rm J} \equiv \hbar I_0/2 e$, with a charging energy
$E_{\rm c} \equiv (2e)^2/2 C \ll E_{\rm J}$. The dynamics of the Josephson phase difference
$\delta$ is that of a particle of mass $M = \hbar^2 C / 4 e^2$ moving in an effective potential
$U(\delta) \equiv - E_{\rm J} (\cos \delta + s \, \delta)$, for bias current $s = I_{\rm b}/I_{0}$
\cite{Ba82,Fulton1974}. For bias currents $s < 1$, the potential $U(\delta)$ has metastable minima,
separated from the continuum by a barrier $\Delta U \equiv U(\delta_{\rm max}) - U(\delta_{\rm
min}) \rightarrow (4\sqrt{2}/3) E_{\rm J} (1-s)^{3/2}$ for $s \rightarrow 1^-$, as shown in Fig.
\ref{fig.cubicpotential}. The curvature $U''(\delta)$ defines the small-amplitude plasma frequency
$\omega_{\rm p} \equiv \sqrt{U''(\delta_{\rm min})/M} = \omega_{{\rm p}0} (1-s^2)^{1/4}$, with
$\omega_{{\rm p}0} = \sqrt{2 e I_0 /\hbar C} = \sqrt{2 E_{\rm c} E_{\rm J}}/\hbar$. The Hamiltonian
for the junction phase difference is $H_{\rm J} = P^2/2M + U(\delta)$, with $P = -i \hbar d/d
\delta$ the momentum operator. The junction's zero-voltage state corresponds to the phase
``particle'' trapped in one of the metastable minima.

The lowest two quasi-bound states in a local minimum, $|0\rangle$ and $|1\rangle$, define the phase
qubit. State preparation is typically carried out with $s$ just below unity, in the range $s =
0.95-0.99$, where $U(\delta)$ is strongly anharmonic, and for which there are only a few quasibound
states \cite{Martinis2002,Wellstood:2003}. The anharmonicity allows state preparation from a
classical radiofrequency (rf) field, as then the frequency of the classical field can be set to
couple to only the lowest two states. In our scheme, by contrast, single quanta are exchanged
between the junction and the resonator, so anharmonicity is not necessary; we find it convenient to
work with $s$ between 0.5 and 0.9.

\begin{figure}
    \includegraphics[width=0.45\textwidth]{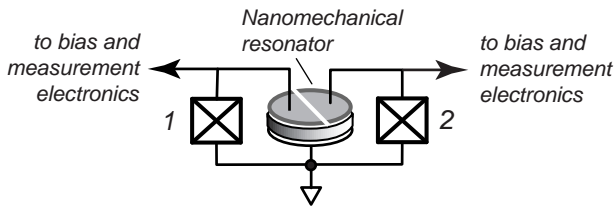}
    \caption[]{\label{fig.twoqubit} Proposed architecture for two-qubit entanglement.
    The qubits are the two Josephson junctions, coupled
    to a resonator, shown as
    a disc with a split gate.}
\end{figure}

We focus here on coupling a single resonator to two Josephson qubits; extensions to larger systems
will be considered in later work. The two-junction circuit is shown in Fig.~\ref{fig.twoqubit}. The
disk-shaped element is the nanomechanical resonator, consisting of a single-crystal piezoelectric
disc sandwiched between two metal plates, and the junctions are the crossed boxes on either side of
the resonator, interrogated by high-impedance circuits \cite{Martinis2002}.

The phase qubit state $|0\rangle$ of a single junction is prepared by waiting for any excited
component to decay. The pure state $|1\rangle$, or a superposition state $\alpha |0\rangle + \beta
|1\rangle$, is prepared by adding a classical rf current $I_{\rm rf}$ to the bias, $I_{\phi 1}(t) =
I_{\rm dc} + I_{\rm rf}^{\rm c} \cos(\omega_{\rm rf}t) + I_{\rm rf}^{\rm s} \sin(\omega_{\rm
rf}t)$. Both $I_{\rm dc}$ and $I_{\rm rf}^{\rm s,c}$ vary slowly compared to $\hbar / \Delta E$.
When $\omega_{\rm rf}$ is near resonance with the level spacing $\Delta E/\hbar$, the qubit will
undergo Rabi oscillations, allowing the controlled preparation of linear combinations of
$|0\rangle$ and $|1\rangle$.

The nanomechanical resonator is designed with a fundamental thickness resonance frequency
$\omega_0/2 \pi \sim 1 - 10 \, {\rm GHz}$, with quality factor $Q \sim 10^5 - 10^6$. Piezoelectric
dilatational resonators with resonance frequencies in this range, and quality factors of $10^3$ at
room temperature, have been fabricated from sputtered AlN \cite{Ruby1994,Ruby2001}. Single-crystal
AlN can also be grown by chemical vapor deposition \cite{Cl01}. Our simulations are based on such a
resonator, with a diameter $d = 1.16~\mu$m and thickness $b = 0.5~\mu m$ \cite{piezoproperties}.
Such resonators can be used to coherently store a qubit state prepared in a current-biased
Josephson junction, return it to that junction, or transfer it to another junction, as well as
entangle two or more junctions. These operations are performed by tuning the energy level spacing
$\Delta E$ into resonance with $\hbar \omega_0$, generating electromechanical Rabi oscillations.

Referring to Fig.~\ref{fig.twoqubit}, the total bias current of junction 1 is $I_{\rm dc1} + I_{\rm
res}$, where $I_{\rm res}$ is the current through the resonator from that junction. A simple model
for the resonator allows us to write $I_{\rm res} = C_{\rm res} (\dot{V} + h_{33} b \dot{U}) $,
where $C_{\rm res}$ is the resonator geometric capacitance, $h_{33}$ the relevant piezoelectric
coupling constant \cite{Au90}, $\dot{V}$ the rate of voltage change, and $\dot{U}$ the rate of
change of the mechanical strain. The current $I_{\rm res}$ is partly due to the capacitance $C_{\rm
res}$ and partly due to the piezoelectrically-coupled strain $U$. $C_{\rm res}$, in parallel with
the junction capacitance $C$, renormalizes the mass $M$ to $\tilde{M} = \hbar^2 \tilde{C} / 4 e^2$,
where $\tilde{C} =C+C_{\rm res}$.

\begin{figure}
\includegraphics[width=.4\textwidth]{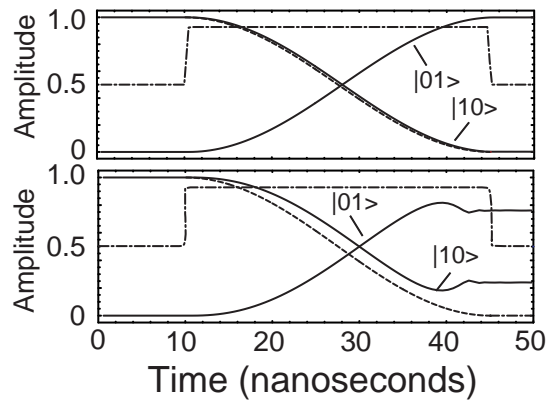}
\caption[]{\label{fig.storage}(a) Phase qubit storage. Solid curves are for $|c_{10}(t)|^2$ and
$|c_{01}(t)|^2$, dashed curve is analytic RWA for $|c_{10}(t)|^2$, dash-dotted curve is $s(t)$. (b)
Qubit storage with arctangent bias-current profile. All other parameters are the same as in (a).
Solid curves are numerical results for $|c_{10}(t)|^2 \!$ and $|c_{01}(t)|^2$.}
\end{figure}

With the resonator coupled to the superconducting phase through the voltage $V$, the Hamiltonian
for the combined junction-resonator system is $H = H_{\rm J}+H_{\rm res} + \delta H$. Here $H_{\rm
res} = \hbar \omega_0 a^\dagger a$ is the Hamiltonian of the isolated resonator, where we have
quantized the resonator displacement field with creation (destruction) operators $a^\dagger$ ($a$),
and only included the fundamental dilatational mode. $\delta H$ is the phase-resonator interaction,
\begin{equation}
\delta H = \frac{\hbar C_{\rm res} b h_{33}}{2 e (1-\eta)}  \delta {\dot U} \, \delta = i g (a -
a^\dagger) \delta,
\end{equation}
where $\eta = 0.054$ and the coupling constant $g$ is
\begin{equation}
g = \frac{ \hbar^{3/2} \, C_{\rm res} h_{33} \, \sqrt{\omega_0} }{ (1 - \eta) e \, \sqrt{\rho \pi b
d^2/4 }}. \label{coupling strength formula}
\end{equation}
For our model resonator $g \approx 0.820~\mu$eV.

\begin{table}
\caption{\label{storage table} Final state amplitudes $c_{mn}(\pi/\Omega_{\rm d})$ for phase-qubit
coupled to nanomechanical resonator.}
\begin{ruledtabular}
\begin{tabular}{|c|ccc|}
probability amplitude & ${\rm Re} \, c_{mn}$ & ${\rm Im} \, c_{mn}$  & $|c_{mn}|^2$ \\ \hline
$c_{00}$ & $0.010$ & $-0.003$ & $0.000$ \\
$c_{01}$ & $-0.257$ & $-0.966$ & $1.000$ \\
$c_{10}$ & $0.009$ & $0.041$ & $0.002$ \\
$c_{11}$ & $-0.010$ & $0.003$ & $0.000$
\end{tabular}
\end{ruledtabular}
\vskip 0.2in
\end{table}

In the junction eigenstate basis, the junction Hamiltonian is $H_{\rm J} = \sum_m \epsilon_m
c_m^{\dagger} c_m$, with creation (destruction) operators $c_m^\dagger$ ($c_m$) acting on the phase
qubit states. The interaction Hamiltonian is
\begin{equation}
\delta H =  i g \sum_{mm'} \langle m | \delta | m' \rangle \, c_m^\dagger c_{m'} (a - a^\dagger).
\label{free hamiltonian}
\end{equation}
The eigenstates of the noninteracting Hamiltonian $H_0 = H_{\rm J}+H_{\rm res}$ are $|mn\rangle
\equiv |m\rangle_{\rm J} \otimes |n\rangle_{\rm res}$, with energies $E_{mn} = \epsilon_m + \hbar
\omega_0 \, n$, where $n$ is the resonator occupation number. An arbitrary state can be expanded as
$|\Psi(t)\rangle = \sum_{mn} c_{mn}(t)|mn\rangle \exp(-i E_{mn} t/\hbar)$.

The full Hamiltonian is equivalent to a few-level atom in an electromagnetic cavity. The cavity
``photons'' are phonons, which interact with the ``atoms'' (here the Josephson junctions) via the
piezoelectric effect. This analogy allows us to adapt quantum-information protocols developed for
cavity-QED to our architecture.

\begin{figure}
    \includegraphics[width=.4\textwidth]{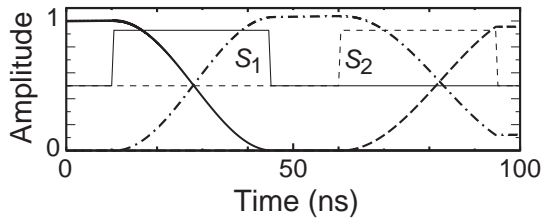}
    \caption[]{\label{fig.transfer}Qubit transfer between two junctions. Solid curve is $|c_{100}(t)|^2 \!$,
    dashed-dotted curve is $|c_{001}(t)|^2 \!$, dashed curve is $|c_{010}(t)|^2 \!$. Thin solid and dotted
    curves show $s_1(t)$ and $s_2(t)$.}
   \vskip 0.2in
\end{figure}

We first show that we can coherently transfer a qubit state from a junction to a resonator, using
the adiabatic approximation combined with the rotating-wave approximation (RWA) of quantum optics
\cite{Scully1997}. We assume that the bias current $s$ changes slowly on the time scale
$\hbar/\Delta E$, and work at temperature $T=0$. The RWA is valid when $\Delta E$ and $\hbar
\omega_0$ are close on the scale of $\hbar \omega_0/Q_{\rm res}$, and when the interaction strength
$g \ll \Delta E$. At time $t=0$, we prepare the resonator in the state $|0\rangle_{\rm res}$. In
the RWA, neglecting relaxation, we obtain the amplitude evolution
\begin{eqnarray}
i \hbar \, \partial_t c_{0n} &=& -i g \, \sqrt{n} \ \langle0|\delta|1\rangle \,
e^{i \omega_{\rm d} t} \, c_{1,n-1} \nonumber \\
i \hbar \, \partial_t c_{1n} &=&  i g \, \sqrt{n+1} \  \langle1|\delta|0\rangle \,
e^{-i \omega_{\rm d} t} \, c_{0,n+1},
\label{RWA equations}
\end{eqnarray}
where $\omega_{\rm d} \equiv  \omega_0 - \Delta E/ \hbar$ is the resonator--qubit detuning. We
integrate to find the reduced density matrices $\rho_{\rm J}(t)$ (in the qubit subspace) and
$\rho_{\rm res}(t)$ (in the zero- and one-phonon resonator subspace). The junction phase is
initially prepared in the pure state $\alpha|0\rangle_{\rm J}+\beta|1\rangle_{\rm J}$,
corresponding to the reduced density matrix
\begin{equation}
\rho_{\rm J}(0) =  \begin{bmatrix} |\alpha|^2 &  \alpha \beta^* \\
\alpha^* \beta & |\beta|^2 \end{bmatrix}.
\end{equation}
We allow the junction and resonator to interact on resonance for a time $\Delta t = \pi/\Omega_d$,
where the Rabi frequency is $\Omega_d = (\Omega_0^2 + \omega_{\rm d}^2)^{1/2}$, in terms of the
tuned (resonant) value $\Omega_0 = 2 g |\langle0|\delta|1\rangle|/\hbar$. After the interval
$\Delta t$, the \emph{resonator} is found to be in the same pure state,
\begin{equation}
\rho_{\rm res}(\pi/\Omega_d) =  \begin{bmatrix} |\alpha|^2
&  - \alpha \beta^* e^{i \pi \omega_0 / \Omega_d} \\
- \alpha^* \! \beta \, e^{-i \pi \omega_0 / \Omega_d} & |\beta|^2 \end{bmatrix},
\end{equation}
apart from expected phase factors. The phase qubit state has been swapped with that of the
resonator. The cavity-QED analog of this operation has been demonstrated experimentally in
Ref.~\cite{Maitre1997}.

\begin{table}
\caption{\label{transfer table} Final amplitudes $c_{m_1 m_2 n}$ for state transfer.}
\begin{ruledtabular}
\begin{tabular}{|c|ccc|}
probability amplitude & ${\rm Re} \, c_{m_1 m_2 n}$ & ${\rm Im} \, c_{m_1 m_2 n}$  & $|c_{m_1 m_2
n}|^2$ \\ \hline
$c_{100}$ & $0.038$ & $-0.013$ & $0.002$ \\
$c_{001}$ & $-0.314$ & $0.152 $ & $0.121$ \\
$c_{010}$ & $-0.882$ & $0.422$ & $0.956$
\end{tabular}
\end{ruledtabular}
\vskip 0.2in
\end{table}

To assess the limitations of the RWA, we also numerically integrated the exact amplitude equations
\begin{equation}
i \hbar \, {\dot c}_{mn} =  \sum_{m' n'} \langle mn| \delta H | m'n' \rangle
e^{i(E_{mn}-E_{m'n'})t/\hbar} \, c_{m'n'}. \label{exact amplitude equations}
\end{equation}
The Josephson junction had parameters corresponding Ref.~\cite{Martinis2002}, $E_{\rm J} = 43.05 \,
{\rm meV}$ and $E_{\rm c} = 53.33 \, {\rm neV}$. We used a 4th-order Runge-Kutta method with a time
step of $10 \, {\rm fs}$. Our main result is shown in Fig.~\ref{fig.storage}. The qubit transfer
depends sensitively on the {\it shape} of the profile $s(t)$, which starts at $s=0.50$, and is then
adiabatically changed to the resonant value $s=0.928$. We find that the time during which $s$
changes should be at least exponentially localized. This can be understood by recalling that the
RWA requires the qubit to be exactly in resonance with the resonator (in the $Q \rightarrow \infty$
limit). Therefore one must bring the system into resonance as quickly as possible without violating
adiabaticity. The power-law tails associated with an arctangent function, for example, lead to
large deviations from the desired behavior, shown in Fig.~\ref{fig.storage}(b).  The result in
Fig.~\ref{fig.storage}(a) was obtained using trapezoidal profiles with a cross-over time of $0.5\,
{\rm ns}$. All quasibound junction states were included in the calculation, and convergence with
the resonator's Hilbert space dimension was obtained. The junction is held in resonance for half a
Rabi period $\pi/\Omega_d$, during which energy is exchanged at the Rabi frequency. The systems are
then brought out of resonance. The final state amplitudes are given in Table~\ref{storage table},
and are quite close to the RWA results.

To pass a qubit state $\alpha |0\rangle + \beta |1\rangle$ from junction 1 to junction 2, the state
is loaded into the first junction and the bias current changed to bring the junction into resonance
with the resonator for half a Rabi period. This writes the state $\alpha |0\rangle + \beta
|1\rangle$ into the resonator. After the first junction is taken out of resonance, the second
junction is brought into resonance for half a Rabi period, passing the state to the second
junction. We have simulated this operation numerically, assuming two identical junctions coupled to
the resonator described above. The results are shown in Fig. \ref{fig.transfer} and Table
\ref{transfer table}, where $c_{m_1 m_2 n}$ is the probability amplitude (in the interaction
representation) to find the system in the state $|m_1 m_2 n \rangle$, with $m_1$ and $m_2$
labelling the states of the two junctions.

\begin{figure}
\includegraphics[width=.4\textwidth]{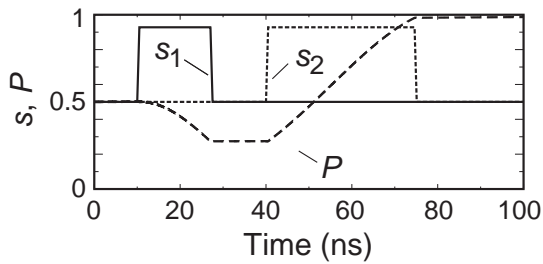}
\caption[]{\label{fig.entangle1}Preparation of entangled Josephson junctions. The solid and dotted
lines are $s_1(t)$ and $s_2(t)$, respectively, and the dashed curve indicates the probability for
the system to be found in the state $(|100\rangle - |010\rangle)/\sqrt{2}$. }
\end{figure}

We can prepare an entangled state of two junctions by bringing the first junction into resonance
with the resonator for 1/4th of a Rabi period \cite{Hagley:1997}, which, according to our RWA
analysis, produces the state $(|100\rangle - |001\rangle)/\sqrt{2}$. After bringing the second
junction into resonance for half a Rabi period, the state of the resonator and second junction are
swapped, leaving the system in the state $(|100\rangle - |010\rangle)/\sqrt{2}$ with a probability
of 0.987, where the resonator is in the ground state and the junctions entangled, as demonstrated
in Fig. \ref{fig.entangle1}. Using the cavity-QED analogy, it will be possible to transfer the
methodology developed for the standard two-qubit operations, in particular controlled-NOT logic, to
this system, using mostly existing technology and demonstrated techniques.

\textbf{Acknowledgements.} ANC and MRG were supported by the Research Corporation, and MRG was by
NSF CAREER Grant No.~DMR-0093217.

\bibliography{qcscheme.bib}

\begin{thebibliography}{30}
\bibitem[{pie()}]{piezoproperties}
\bibinfo{note}{AlN has a density $\rho = 3.26$ g/cm$^3$, piezoelectric modulus
  $e_{33}$ = 1.46 C/m$^2$, and dielectric constant $\epsilon$ = 10.7
  $\epsilon_0$. The resonator has a geometric capacitance $C_{\rm res} = 0.20$
  fF and a resonance frequency $\omega_0/2\pi =$ 10 GHz.}
\end{thebibliography}

\end{document}